\documentclass{ws-procs9x6}

\usepackage[normalem]{ulem} 
\usepackage[dvips]{color} 
\renewcommand\sout{\bgroup \color{red} \ULdepth=-.5ex \ULset}

\begin{document}

\title{CONSTRAINING THE SYMMETRY ENERGY FROM THE NEUTRON SKIN THICKNESS OF TIN ISOTOPES}

\author{LIE-WEN CHEN}

\address{Department of Physics, Shanghai Jiao Tong University, Shanghai 200240, China\\
E-mail: lwchen@sjtu.edu.cn}

\author{CHE MING KO$^*$, JUN XU$^{\dag}$}

\address{Cyclotron Institute and Department of Physics and Astronomy, Texas A\&M
University, College Station, Texas 77843-3366, USA\\
$^*$E-mail: ko@comp.tamu.edu; $^{\dag}$E-mail: xujun@comp.tamu.edu}

\author{BAO-AN LI}

\address{Department of Physics and Astronomy, Texas A\&M University-Commerce,
Commerce, Texas 75429-3011, USA\\
E-mail: Bao-An\_Li@tamu-commerce.edu}

\begin{abstract}
We show in the Skyrme-Hartree-Fock approach that unambiguous
correlations exist between observables of finite nuclei and nuclear
matter properties. Using this correlation analysis to existing data
on the neutron skin thickness of Sn isotopes, we find important
constraints on the value $E_{\text{\textrm{sym}}}({\rho _{0}})$ and
density slope $L$ of the nuclear symmetry energy at saturation
density. Combining these constraints with those from recent analyses
of isospin diffusion and double neutron/proton ratio in heavy ion
collisions leads to a value of $L=58\pm 18$ MeV approximately
independent of $E_{\text{\textrm{sym}}}({\rho _{0}})$.
\end{abstract}


\bodymatter

\section{Introduction}\label{aba:sec1}

The nuclear symmetry energy $E_{\text{\textrm{sym}}}(\rho )$ plays a
crucial role in both nuclear physics and astrophysics
\cite{Ste05,LCK08}. Although significant progress has been made in
recent years in determining the density dependence of
$E_{\text{\textrm{sym}}}(\rho )$ \cite{LCK08}, large uncertainties
still exist even around the normal density $\rho _{0}$ \cite{XuC10},
and this has hindered us from understanding more precisely many
important properties of neutron stars \cite{XuJ09}. To constrain the
symmetry energy with higher accuracy is thus of crucial importance.

Theoretically, it has been established \cite%
{Bro00,Hor01,Fur02,Yos04,Che05b,Tod05,Rei10,Dan03,Cen09} that the
neutron skin thickness $\Delta r_{np}=\langle r_{n}^{2}\rangle
^{1/2}-\langle r_{p}^{2}\rangle ^{1/2} $ of heavy nuclei, given by
the difference of their neutron and proton
root-mean-squared radii, provides a good probe of $E_{\text{\textrm{sym}}%
}(\rho )$. In particular, $\Delta r_{np}$ has been found to
correlate strongly with both $E_{\text{\textrm{sym}}}(\rho_0 )$ and
$L$ in microscopic mean-field calculations
\cite{Bro00,Hor01,Fur02,Yos04,Che05b,Tod05,Rei10}. It is, however,
difficult to extract an accurate value for $L $ from comparing the
calculated $\Delta r_{np}$ of heavy nuclei with experimental data as
it depends on several nuclear interaction parameters in a highly
correlated manner \cite{Fur02,Rei10} and the calculations have been
usually carried out by varying the interaction parameters. A
well-known example is the Skyrme-Hartree-Fock (SHF) approach using
normally $9$ interaction parameters and there are more than $120$
sets of Skyrme interaction parameters in the literature.

In the present talk, we report our recent work \cite{Che10} on a new
method to analyze the correlation between observables of finite
nuclei and some macroscopic properties of asymmetric nuclear matter.
Instead of varying directly the $9$ interaction parameters within
the SHF, we express them explicitly in terms of $9$ macroscopic
observables that are either experimentally well constrained or
empirically well known. Then, by varying individually these
macroscopic observables within their known ranges, we can examine
more transparently the correlation of $\Delta r_{np}$ with each
individual observable. In particular, we have demonstrated that
important constraints on $E_{\text{\textrm{sym}}}({\rho _{0}})$ and
$L$ can be obtained with the application of this correlation
analysis to existing data on the neutron skin thickness of Sn
isotopes.

\section{The theoretical method}

In the standard SHF model \cite{Cha97}, the $9$ Skyrme interaction
parameters, i.e., $\sigma $, $t_{0}-t_{3}$, $x_{0}-x_{3}$ can be
expressed analytically in terms of $9$ macroscopic quantities $\rho
_{0}$, $E_{0}(\rho _{0})$, the incompressibility $K_{0}$, the
isoscalar effective mass $m_{s,0}^{\ast }$,
the isovector effective mass $m_{v,0}^{\ast }$, $E_{\text{\textrm{sym}}}({%
\rho _{0}})$, $L$, the gradient coefficient $G_{S}$, and the
symmetry-gradient coefficient $G_{V}$~\cite{Che10,Che11}, i.e.,
\begin{eqnarray}
\sigma &=&\gamma -1,\quad t{_{0}}=4\alpha /(3{\rho _{0}}), \quad x{_{0}}
=3(y-1)E_{\text{\textrm{sym}}}^{\mathrm{loc}}({\rho _{0}})/\alpha-1/2, \notag \\
t_{1} &=&20C/\left[ 9{\rho _{0}(}k_{\mathrm{F}}^{0})^{2}\right]
+8G_{S}/3,\quad x_{1}=\left[ 12G_{V}-4G_{S}-\frac{6D}{{\rho _{0}(}k_{\mathrm{F}}^{0})^{2}}%
\right] /(3t_{1}),\notag\\
t_{2} &=&\frac{4(25C-18D)}{9{\rho _{0}(}k_{\mathrm{F}}^{0})^{2}}-\frac{%
8(G_{S}+2G_{V})}{3},\notag\\
x_{2} &=&\left[ 20G_{V}+4G_{S}-\frac{5(16C-18D)}{3{\rho _{0}(}k_{\mathrm{F}%
}^{0})^{2}}\right] /(3t_{2}), \notag\\
t{_{3}} &=&16\beta /\left[ {\rho _{0}}^{\gamma }(\gamma +1)\right],\quad
x{_{3}} =-3y(\gamma +1)E_{\text{\textrm{sym}}}^{\mathrm{loc}}({\rho _{0}}%
)/(2\beta )-1/2,
\end{eqnarray}%
where $k_{\mathrm{F}}^{0}=(1.5\pi ^{2}{\rho _{0}})^{1/3}$,
$E_{\text{\textrm{sym}}}^{\mathrm{loc}}({\rho _{0}})=E_{\text{\textrm{sym}}%
}({\rho
_{0}})-E_{\text{\textrm{sym}}}^{\mathrm{kin}}({\rho_{0}})-D$, and
the parameters $C$, $D$, $\alpha $, $\beta $, $\gamma $, and $y$ are
defined as \cite{Che09}
\begin{eqnarray}
C&=&\frac{m-m_{s,0}^{\ast }}{m_{s,0}^{\ast }}E_{\mathrm{kin}}^{0},\quad
D =\frac{5}{9}E_{\mathrm{kin}}^{0}\left( 4\frac{m}{m_{s,0}^{\ast }}-3\frac{%
m}{m_{v,0}^{\ast }}-1\right), \notag \\
\alpha &=&-\frac{4}{3}E_{\mathrm{kin}}^{0}-\frac{10}{3}C-\frac{2}{3}[E_{%
\mathrm{kin}}^{0}-3E_{0}(\rho _{0})-2C]\notag\\
&&\times\frac{K_{0}+2E_{\mathrm{kin}}^{0}-10C}{K_{0}+9E_{0}(\rho _{0})-E_{%
\mathrm{kin}}^{0}-4C},\notag \\
\beta &=&\left[\frac{E_{\mathrm{kin}}^{0}}{3}-E_{0}(\rho
_{0})-\frac{2}{3}C\right]\frac{K_{0}-9E_{0}(\rho _{0})+5E_{\mathrm{kin}}^{0}-16C}{%
K_{0}+9E_{0}(\rho _{0})-E_{\mathrm{kin}}^{0}-4C}, \notag \\
\gamma &=&\frac{K_{0}+2E_{\mathrm{kin}}^{0}-10C}{3E_{\mathrm{kin}%
}^{0}-9E_{0}(\rho _{0})-6C},\quad
y =\frac{L-3E_{\text{\textrm{sym}}}({\rho _{0}})+E_{\text{\textrm{\ sym}}%
}^{\mathrm{kin}}({\rho _{0}})-2D}{3(\gamma -1)E_{\text{ \textrm{sym}}}^{%
\mathrm{loc}}({\rho _{0}})},
\end{eqnarray}%
with $E_{\mathrm{kin}}^{0}=\frac{3\hbar ^{2}}{10m}\left(
\frac{3\pi^{2}}{2}\right) ^{2/3}\rho _{0}^{2/3}$ and
$E_{\text{\textrm{sym}}}^{\rm kin}({\rho _{0}})= \frac{\hbar
^{2}}{6m}\left( \frac{3\pi ^{2}}{2}{\rho _{0}}\right) ^{2/3}$.

As a reference for the correlation analyses below, we use the MSL0
parameter set~\cite{Che10}, which is obtained by using the following
empirical values for the macroscopic quantities: $\rho
_{0}=0.16$ fm$^{-3}$, $E_{0}(\rho _{0})=-16$ MeV, $K_{0}=230$ MeV, $%
m_{s,0}^{\ast }=0.8m $, $m_{v,0}^{\ast }=0.7m$, $E_{\text{\textrm{sym}}}({%
\rho _{0}})=30$ MeV, and $L=60$ MeV, $G_{V}=5$ MeV$\cdot $fm$^{5}$, and $%
G_{S}=132$ MeV$\cdot $fm$^{5}$. And the spin-orbital coupling
constant $W_{0}=133.3$ MeV $\cdot $fm$^{5}$ is used to fit the
neutron $p_{1/2}-p_{3/2}$ splitting in $^{16}$O. Using other Skyrme
interactions obtained from fitting measured binding energies and
charge rms radii of finite nuclei does not change our conclusion.

\section{Results}

\begin{figure}[tbp]
\begin{center}
\includegraphics[scale=0.9]{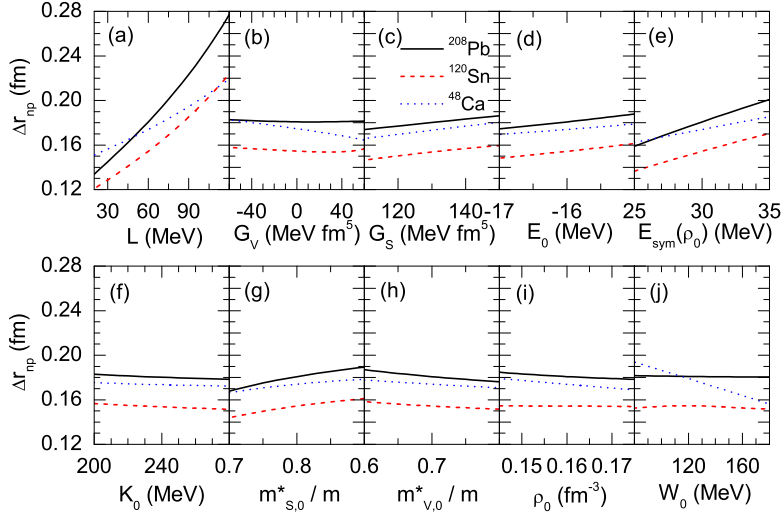}
\end{center}
\caption{(Color online) The neutron skin thickness $\Delta r_{np}$
of $^{208}$ Pb, $^{120}$Sn and $^{48}$Ca from SHF with MSL0 by
varying individually $L$ (a), $G_{V}$ (b), $G_{S}$ (c), $E_{0}(\rho
_{0})$ (d), $E_{\text{\textrm{sym}}}(\rho _{0})$ (e), $K_{0}$ (f),
$m_{s,0}^{\ast }$ (g), $m_{v,0}^{\ast }$ (h), $\rho _{0}$ (i), and
$W_{0}$ (j). Taken from Ref. \cite{Che10}.}
\label{RnpPbSnCa}
\end{figure}

To reveal clearly the dependence of $\Delta r_{np}$ on each
macroscopic quantity, we vary one quantity at a time while keeping
all others at their
default values in MSL0. Shown in Fig. \ref{RnpPbSnCa} are the values of $%
\Delta r_{np}$ for $^{208}$Pb, $^{120}$Sn and $^{48}$Ca. Within the
uncertain ranges for the macroscopic quantities considered here, the
$\Delta r_{np}$ of $^{208}$Pb and $^{120}$Sn exhibits a very strong
correlation with $L$. However, it depends
only moderately on $E_{\text{\textrm{sym}}}({\rho _{0}})$ and weakly on $%
m_{s,0}^{\ast }$. On the other hand, the $\Delta r_{np}$ of $^{48}$Ca
displays a much weaker dependence on both $L$ and $E_{\text{\textrm{sym}}}({%
\rho _{0}})$. Instead, it depends moderately on $G_{V}$ and $W_{0}$.
This explains the weaker $\Delta
r_{np}$-$E_{\text{\textrm{sym}}}({\rho })$ correlation observed for
$^{48}$Ca in previous SHF calculations using different interaction
parameters \cite{Che05b}.

\begin{figure}[htb]
\begin{minipage}{14.8pc}
\includegraphics[scale=0.8]{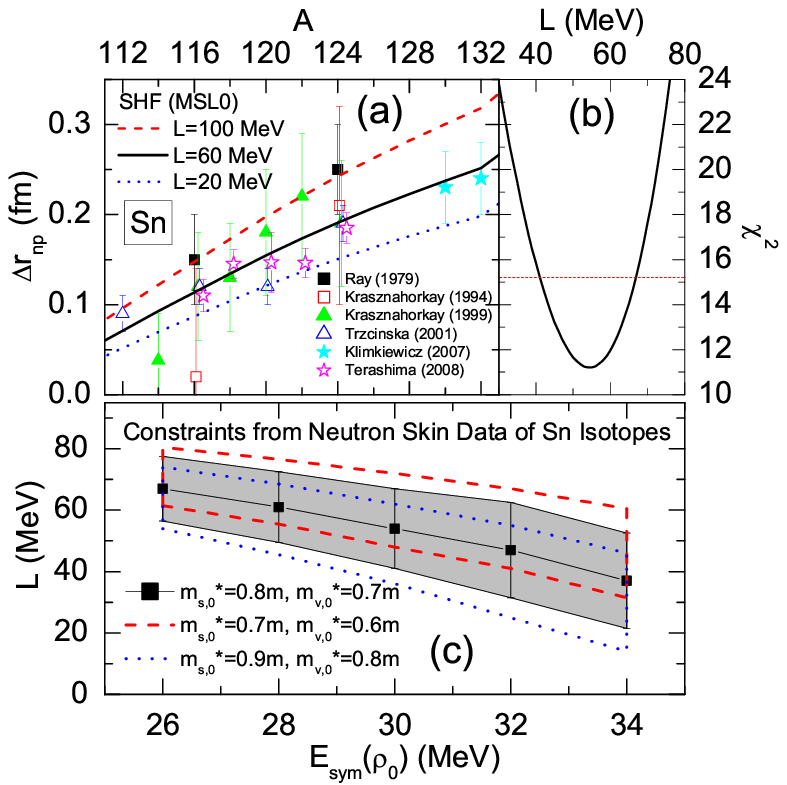}
\end{minipage}
\begin{minipage}{12.0pc}
\includegraphics[scale=0.64]{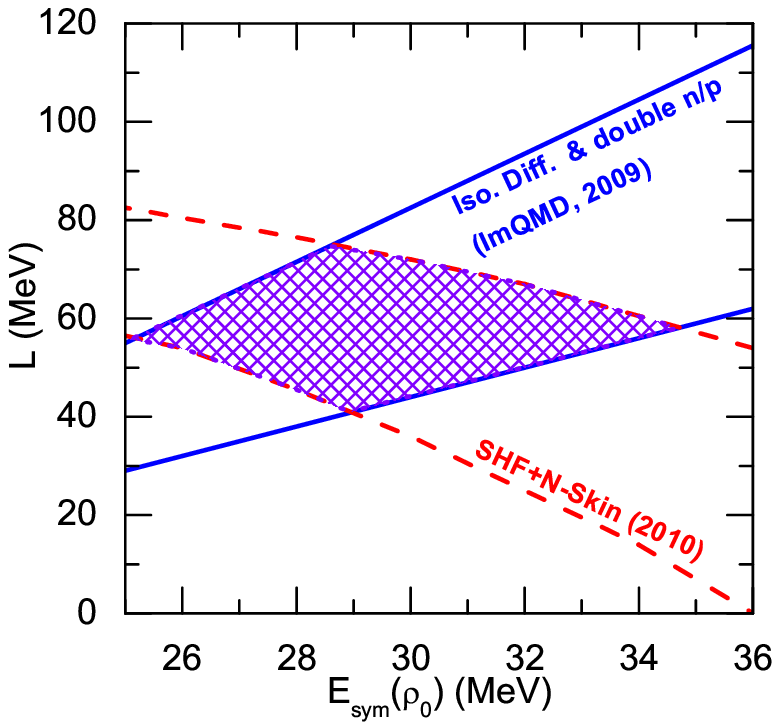}
\end{minipage}
\caption{(Color online) Left window: (a) The $\Delta r_{np}$ data
for Sn isotopes from different experimental methods and results from
SHF calculation using MSL0 with $L=20$, $60$ and $100$ MeV. (b)
$\chi ^{2}$ as a function of $L$. (c) Constraints on $L$ and
$E_{\text{\textrm{sym}}}(\rho _{0})$ from the $\chi ^{2}$ analysis
of the $\Delta r_{np}$ data on Sn isotopes (Grey band as well as
dashed and dotted lines). Right window: Constraints on $L$ and
$E_{\text{\textrm{sym}}}(\rho _{0})$ obtained in the present work
(dashed lines) and that from Ref. \cite{Tsa09} (solid lines). The
shaded region represents their overlap. Taken from Ref.
\cite{Che10}.} \label{RnpSnL}
\end{figure}

Experimentally, the $\Delta r_{np}$ of heavy Sn isotopes has been
systematically measured \cite{Ray79,Kra94,Kra99,Trz01,Kli07,Ter08}.
As an illustration, we first show in the panel (a) of left window in
Fig. \ref{RnpSnL} the comparison of the available Sn $\Delta r_{np}$
data with our calculated results using $20 $, $60$ and $100$ MeV,
respectively, for the value of $L$ and the default values for all
other quantities in MSL0. It is seen that the value $L=60$ MeV best
describes the data. To be more precise, the $\chi ^{2}$ evaluated
from the difference between the theoretical and experimental $\Delta
r_{np}$ values is shown as a function of $L$ in the panel (b) of
left window in Fig. \ref{RnpSnL}. The most reliable value of $L$ is
found to be $L=54\pm 13$ MeV within a $2\sigma $ uncertainty.

Since the value of $\Delta r_{np}$ depends on both $L$ and $E_{\text{\textrm{%
sym}}}({\rho _{0}})$, we have carried out a two-dimensional $\chi
^{2}$ analysis as shown by the grey band in the panel (c) of left
window in Fig. \ref{RnpSnL}. It is seen that increasing the value of
$E_{\text{\textrm{sym}}}({\rho _{0}})$ systematically leads to
smaller values of $L$. Furthermore, we have estimated the effects of
nucleon effective mass by using $m_{s,0}^{\ast }=0.7m$ and
$m_{v,0}^{\ast }=0.6m$ as well as $m_{s,0}^{\ast
}=0.9m$ and $m_{v,0}^{\ast }=0.8m$, in accord with the empirical constraint $%
m_{s,0}^{\ast }>m_{v,0}^{\ast }$ \cite{LCK08,XuC10,Les06}, and the
resulting constraints are shown by the dashed and dotted lines. As
expected from the results shown in Fig.\ \ref{RnpPbSnCa}, effects of
nucleon effective mass are small with the value of $L$ shifting by
only a few MeV for a given $E_{\text{\textrm{sym}}}( {\rho _{0}})$.
As one has also expected, effects of varying other macroscopic
quantities are even smaller.

The above constraints on the $L$-$E_{\text{\textrm{sym}}}({\rho
_{0}})$ correlation can be combined with those from recent analyses
of isospin diffusion and double $n/p$ ratio in heavy ion collisions
at intermediate energies \cite{Tsa09} to determine simultaneously
the values of both $L$ and $E_{\text{\textrm{sym}}}({\rho _{0}})$.
Shown in the right window of Fig. \ref{RnpSnL} are the two
constraints in the $E_{\text{\textrm{sym}}}({\rho _{0}})$-$L$ plane.
Interestingly, these two constraints display opposite
$L$-$E_{\text{\textrm{sym}}}({\rho _{0}})$ correlations. This allows
us to extract a value of $L=58\pm 18$ MeV approximately independent
of the value of $E_{\text{\textrm{sym}}}({\rho _{0}})$. This value
of $L$ is quite precise compared to existing estimates in the
literature (See Ref. \cite{XuC10} for a recent summary) although the
constraint on $E_{\text{\textrm{sym}}}({\rho _{0}})$ is not
improved.

\section{Summary}
We have proposed a new method to analyze the correlations between
observables of finite nuclei and some macroscopic properties of
nuclear matter, and demonstrated that the existing neutron skin data
on Sn isotopes can give important constraints on the symmetry energy
parameters $L$ and $E_{\text{\textrm{sym}}}({\rho _{0}})$. Combining
these constraints with those from recent analyses of isospin
diffusion and double $n/p$ ratio in heavy ion collisions leads to a
quite accurate value of $L=58\pm 18$ MeV approximately independent
of $E_{\text{\textrm{sym}}}({\rho _{0}})$.

\section*{Acknowledgments}
This work was supported in part by the NNSF of China under Grant No.
10975097, the National Basic Research Program of China (973 Program)
under Contract No. 2007CB815004, U.S. NSF under Grant No.
PHY-0758115 and PHY-0757839, the Welch Foundation under Grant No.
A-1358, the Research Corporation under Award No. 7123, the Texas
Coordinating Board of Higher Education Award No. 003565-0004-2007.

\end{document}